\documentclass{emulateapj}

\usepackage{amssymb}

\def\ts     {\thinspace}
\def\kms    {\ts km\ts s$^{-1}$}
\def\etal   {{\rm et\ts al.}}
\def\msol   {M$_{\odot}$}

\def\ci     {C\ts {\scriptsize I}}

\def\aco    {$^{12}${\rm CO}($J$=1$\to$0)}

\def\cco    {$^{12}${\rm CO}($J$=3$\to$2)}

\def\gco    {$^{12}${\rm CO}($J$=7$\to$6)}

\def\ahcn    {{\rm HCN}($J$=1$\to$0)}

\def\ahco    {{\rm HCO$^+$}($J$=1$\to$0)}
\def\dhco    {{\rm HCO$^+$}($J$=4$\to$3)}

\slugcomment{draft version \today, accepted for publication in the Astrophysical Journal Letters}

\shorttitle{HCO+(1-0) in the Cloverleaf}
\shortauthors{Riechers et al.}

\begin{document}

\title{First detection of HCO$^+$ emission at high redshift}

\author{Dominik A. Riechers\altaffilmark{1}, Fabian Walter\altaffilmark{1}, 
Christopher L. Carilli\altaffilmark{2}, Axel Weiss\altaffilmark{3}, \\
Frank Bertoldi\altaffilmark{4}, Karl M. Menten\altaffilmark{3}, 
Kirsten K. Knudsen\altaffilmark{1}, and Pierre Cox\altaffilmark{5}}

\altaffiltext{1}{Max-Planck-Institut f\"ur Astronomie, K\"onigstuhl 17, 
Heidelberg, D-69117, Germany}

\altaffiltext{2}{National Radio Astronomy Observatory, PO Box O,
  Socorro, NM 87801, USA}

\altaffiltext{3}{Max-Planck-Institut f\"ur Radioastronomie, Auf dem
  H\"ugel 69, Bonn, D-53121, Germany}

\altaffiltext{4}{Argelander-Institut f\"ur Astronomie, Universit\"at
  Bonn, Auf dem H\"ugel 71, Bonn, D-53121, Germany}

\altaffiltext{5}{Institut de RadioAstronomie Millim\'etrique, 300 Rue
  de la Piscine, Domaine Universitaire, 38406 Saint Martin d'H\'eres,
  France}

\email{riechers@mpia.de}

\begin{abstract}
  We report the detection of \ahco\ emission towards the Cloverleaf
  quasar ($z=2.56$) through observations with the Very Large Array.
  This is the first detection of ionized molecular gas emission at
  high redshift ($z>2$).  HCO$^+$ emission is a star formation
  indicator similar to HCN, tracing dense molecular hydrogen gas
  ($n({\rm H_2}) \simeq 10^5\,$cm$^{-3}$) within star--forming
  molecular clouds.  We derive a lensing--corrected HCO$^+$ line
  luminosity of $L'_{\rm HCO^+} = 3.5 \times 10^{9}\,$K \kms pc$^2$.
  Combining our new results with CO and HCN measurements from the
  literature, we find a HCO$^+$/CO luminosity ratio of 0.08 and a
  HCO$^+$/HCN luminosity ratio of 0.8. These ratios fall within the
  scatter of the same relationships found for low--$z$ star--forming
  galaxies.  However, a HCO$^+$/HCN luminosity ratio close to unity
  would not be expected for the Cloverleaf if the recently suggested
  relation between this ratio and the far--infrared luminosity were to
  hold.  We conclude that a ratio between HCO$^+$ and HCN luminosity
  close to 1 is likely due to the fact that the emission from both
  lines is optically thick and thermalized and emerges from dense
  regions of similar volumes.  The CO, HCN and HCO$^+$ luminosities
  suggest that the Cloverleaf is a composite AGN--starburst system, in
  agreement with the previous finding that about 20\% of the total
  infrared luminosity in this system results from dust heated by star
  formation rather than heating by the AGN.  We conclude that HCO$^+$
  is potentially a good tracer for dense molecular gas at high
  redshift.
\end{abstract}


\section{Introduction}

One important goal in studies of galaxy formation is to determine star
formation characteristics through observations of molecular gas in the
early universe.  Molecular gas in high--redshift galaxies is commonly
traced by CO emission and has been found in $>$30 galaxies at $z>2$ to
date.  The observed molecular gas with masses of $\geq
10^{10}$\,\msol\ provides the requisite material for star formation
(see review by Solomon \& Vanden Bout \citeyear{sv05}, and references
therein).

CO is a good indicator for the total molecular gas content of a
system, as it can be excited at relatively low densities; its low
dipole moment of $\mu_{\rm D}^{\rm CO} = 0.1$ implies a critical
density of only $n_{\rm H_2} \sim 10^2-10^3\,$cm$^{-3}$ for the
lower--$J$ transitions. Hence, emission from low--$J$ CO transitions
is a relatively poor tracer of the denser gas that is more intimately
associated with star formation. The most common tracers of the dense
molecular gas phase are HCN and HCO$^+$. Both molecules have much
higher dipole moments ($\mu_{\rm D}^{\rm HCN} = 2.98$, $\mu_{\rm
  D}^{\rm HCO^+} = 4.48$) than CO. The critical density to
collisionally thermalize their lower--$J$ transitions is therefore
much higher than for CO, $n_{\rm H_2} \sim 10^5-10^6$\,cm$^{-3}$
(e.g.\ Gao \& Solomon \citeyear{gao04a}, \citeyear{gao04b}; Brouillet
et al.\ \citeyear{bro05}).

Recent studies of the dense molecular gas phase in nearby ($z<0.3$)
luminous and ultra--luminous infrared galaxies (LIRGs/ULIRGs) have
shown that the HCN and HCO$^+$ luminosities correlate with the
star--formation rate as traced by the far--infrared (FIR) luminosity.
These correlations are tighter than the correlation between the CO and
FIR luminosities (Solomon \etal\ \citeyear{sol92a}; Gao \& Solomon
\citeyear{gao04a}, \citeyear{gao04b}; Graci\'a--Carpio et al.\
\citeyear{gra06}).  Due to the relative faintness of the emission
lines, HCN has only been detected in four objects at high $z$ to date,
and all detections are in the host galaxies of quasars (Solomon \etal\
\citeyear{sol03}; Vanden Bout \etal\ \citeyear{vdb04}; Carilli \etal\
\citeyear{car04}; Wagg \etal\ \citeyear{wag05}).

It has recently been argued that AGN--dominated galaxies have higher
HCN/HCO$^+$ and HCN/CO luminosity ratios than starburst--dominated
galaxies (Kohno et al.\ \citeyear{koh01}; Kohno \citeyear{koh05};
Imanishi et al.\ \citeyear{ima06}).  In this context, it has been
suggested that the presence of X--ray emission emerging from a
dust--enshrouded AGN may significantly enhance the chemical abundance
of HCN relative to HCO$^+$ (Lepp \& Dalgarno \citeyear{lep96}; Kohno
et al.\ \citeyear{koh01}; Usero et al.\ \citeyear{use04}).  Also, the
excitation of the HCN molecule may be affected by IR--pumping through
a 14\,$\mu$m vibrational band (Aalto et al.\ \citeyear{aal95}).  Based
on these considerations and their HCO$^+$ survey of low--$z$ (U)LIRGs,
Graci\'a--Carpio et al.\ (\citeyear{gra06}) suggest that the
HCO$^+$--to--HCN intensity ratio towards FIR--bright ($L_{\rm FIR} >
10^{12}\,$L$_{\odot}$) objects (like the Cloverleaf) are likely low.
Measurements of HCO$^+$ emission in high redshift objects would thus
lead to new constraints on their dense interstellar medium and,
potentially, on the radiation field pervading it.

In this letter, we report the first high--$z$ detection of \ahco\
emission, which was observed towards the Cloverleaf quasar ($z=2.56$)
with the Very Large Array (VLA)\footnote{The Very Large Array is a
  facility of the National Radio Astronomy Observatory, operated by
  Associated Universities, Inc., under a cooperative agreement with
  the National Science Foundation.}.  Due to its strong gravitational
magnification (magnification factor $\mu_{\rm L} = 11$, Venturini \&
Solomon \citeyear{ven03}), the Cloverleaf is the brightest CO source
at high redshift, and it also exhibits bright \ahcn\ emission (Solomon
\etal\ \citeyear{sol03}) and emission from both \ci\ fine structure
lines (Weiss \etal\ \citeyear{wei03}, \citeyear{wei05}). A previous
search for \dhco\ emission in this source has been unsuccessful,
setting an upper limit of 14\,mJy on the peak line flux density
(Wilner \etal\ \citeyear{wil95}).  We use a standard concordance
cosmology throughout, with $H_0 = 73\,$\kms\,Mpc$^{-1}$, $\Omega_{\rm
  M} =0.24$, and $\Omega_{\Lambda} = 0.72$ (Spergel \etal\
\citeyear{spe03}, \citeyear{spe06}).

\section{Observations}

We observed the \ahco\ transition line ($\nu_{\rm rest} =
89.1885230\,$GHz) towards H1413+117 (the Cloverleaf quasar) using the
VLA in D configuration between 2005 November 26 and 2006 January 13.
At the target $z$ of 2.55784, the line is redshifted to 25.068166\,GHz
(11.96\,mm).  The total integration time amounts to 25.5\,h.
Observations were performed in fast--switching mode using the nearby
source 14160+13204 (distance to the Cloverleaf: $1.8^\circ$) for
secondary amplitude and phase calibration. Observations were carried
out under excellent weather conditions with 23 antennas. The phase
stability in all runs was excellent (typically $<$15$^\circ$ rms for
the longest baselines). For primary flux calibration, 3C286 was
observed during each run.

Two 25\,MHz wide intermediate frequency bands (IFs) with seven
3.125\,MHz channels each were observed simultaneously in the
so--called 'spectral line' mode centered at the \ahco\ line frequency,
leading to an effective bandwidth of 43.75\,MHz (corresponding to
523\kms\ at 25.1\,GHz).  In addition, two 50\,MHz (corresponding to
598\kms\ at 25.1\,GHz) IFs were observed in the so--called
'quasi--continuum' mode at 24.7351\,GHz and 24.7851\,GHz to monitor
the continuum at 12\,mm.  We chose a setting with both continuum
channels below the \ahco\ line frequency to avoid the significantly
worse system temperatures and locking problems of the local oscillator
(LO) above 25\,GHz.  The continuum was observed for one third of the
total time to attain the same rms as in the combined line channels.

For data reduction and analysis, the ${\mathcal{AIPS}}$ package was
used. The two continuum channels were concatenated in the
uv/visibility plane.  The \ahco\ line data cube was generated by
subtracting a CLEAN component model of the continuum emission from the
visibility data. All data were mapped using the CLEAN algorithm and
'natural' weighting without applying a further taper; this results in
a synthesized beam of 4.0\,$''$$\times$3.0\,$''$ ($\sim$28\,kpc at $z
= 2.56$).  The final rms in the combined map is 16\,$\mu$Jy
beam$^{-1}$ for a 34.375\,MHz (corresponding to 411\,\kms) channel,
and 50\,$\mu$Jy beam$^{-1}$ for a 6.25\,MHz (75\,\kms) channel.

\section{Results}

We have detected emission from the \ahco\ transition line towards the
Cloverleaf quasar ($z=2.56$).  The source appears to be marginally
resolved in both the continuum and the HCO$^+$ line maps.
Two--dimensional Gaussian fitting yields a continuum peak flux density
of 343 $\pm$ 12 \,$\mu$Jy beam$^{-1}$.  The continuum--subtracted
\ahco\ line map is shown in Fig.\ 1.  The cross indicates the
geometrical center position of the resolved \gco\ map at
$\alpha=14^{\rm h}15^{\rm m}46^{\rm s}.233$,
$\delta=+11^\circ29'43''.50$ (Alloin \etal\ \citeyear{all97}).  The
small offset is likely due to the fact that the lens images are not
equally bright, i.e.\ the center of intensity is offset from the
geometrical center position. The deconvolved source size from the
Gaussian fit is in good agreement with the size of the resolved
structure seen in \gco.  The line is clearly detected at 8$\sigma$
over a range of 34.375\,MHz (411\kms).  We derive a line peak flux
density of 193 $\pm$ 28 \,$\mu$Jy beam$^{-1}$.  In Fig.\ 2, four
channel maps (6.25\,MHz, or 75\,\kms\ each) of the central 25\,MHz
(300\,\kms) of the \ahco\ line are shown.  At an rms of
50\,$\mu$Jy\,beam$^{-1}$, the line is detected at 4$\sigma$ in the
central two channels, and the decline of the line intensity towards
the line wings is clearly visible in the outer channels, as expected.
We attribute the small offset between the peak positions of channels 2
and 3 to observational uncertainties rather than to a real velocity
gradient.  We thus derive a \ahco\ line luminosity of $L'_{\rm HCO^+}
= 3.5 \times 10^9\,$K\,\kms\,pc$^2$ (corrected for gravitational
magnification, $\mu_{\rm L} = 11$, and the finite source size relative
to the synthesized beam, which leads to a 30\% correction based on the
source extension seen in CO, see Table \ref{tab-1} and its caption for
details).

We summarize our results in Table 1 together with the line fluxes and
luminosities for HCN (Solomon et al.\ \citeyear{sol03}) and CO (Weiss
et al.\ \citeyear{wei03}).  Our \ahco\ peak flux density corresponds
to $\sim$80\% of the \ahcn\ peak flux of 0.24 $\pm$ 0.04\,mJy derived
by Solomon et al.\ (\citeyear{sol03}). This corresponds to a $L'_{\rm
  HCO^+}$/$L'_{\rm HCN}$ luminosity ratio of 0.8, which is consistent
with unity within the statistical and systematical uncertainties.
They use an extrapolated continuum peak flux at 24.9\,GHz of $S_{\rm
  cont}(24\,{\rm GHz}) = 0.26 \pm 0.03$\,mJy. The difference to our
value may be due to problems with their extrapolation or calibration
errors. However, it is also possible that the continuum of the
Cloverleaf is variable at 25\,GHz.

\section{Discussion}

In the following, we discuss relationships between the emission
observed in HCO$^+$ and other molecules and the far--IR continuum for
a sample of low--$z$ spiral and starburst galaxies (Nguyen-Q-Rieu et
al.\ \citeyear{ngu92}; Imanishi et al.\ \citeyear{ima04}; Gao \&
Solomon \citeyear{gao04a}), low--$z$ (U)LIRGs (Graci\'a--Carpio et
al.\ \citeyear{gra06}; Imanishi et al.\ \citeyear{ima06}), and the
Cloverleaf (this work; Solomon et al.\ \citeyear{sol03}; Weiss et al.\
\citeyear{wei03}) as shown in Fig.~3. As only an upper limit exists
for the \aco\ line emission in the Cloverleaf (Tsuboi et al.\
\citeyear{tsu99}), we here assume that CO is fully thermalized up to
the 3$\to$2 transition (i.e., $L'_{\rm CO(1-0)} = L'_{\rm CO(3-2)}$).
We do not discuss effects of differential lensing, which could distort
the intrinsic luminosity ratios, as models indicate similar sizes for
molecular and dust emission in the Cloverleaf (Solomon et al.\
\citeyear{sol03}).

{\em Figure 3a}: $L'_{\rm HCO^+}$ correlates closely with $L_{\rm
  FIR}$; a linear least squares fit (excluding the Cloverleaf) yields
log$(L_{\rm FIR}) = (1.11 \pm 0.06) \times {\rm log} (L'_{\rm
  HCO^+})+2.2$. For HCN, Gao \& Solomon (\citeyear{gao04a}) find
log$(L_{\rm FIR}) = 0.97 \times {\rm log} (L'_{\rm HCN})+3.1$ based on
a larger sample of local starburst and spiral galaxies. Both slopes
are consistent with unity.

{\em Figure 3b}: $L'_{\rm HCO^+}$ also correlates closely with
$L'_{\rm HCN}$; a linear least squares fit (again excluding the
Cloverleaf) yields log$(L'_{\rm HCN}) = (0.94 \pm 0.06) \times {\rm
  log} (L'_{\rm HCO^+})+0.6$.  Panels {\bf a} and {\bf b} of Fig.~3
thus exemplify that HCO$^+$ traces dense molecular gas as well as HCN,
and its close correlation with the far--IR continuum emission suggests
that HCO$^+$ may also be used as a star formation indicator.  It is
remarkable how well the Cloverleaf agrees with the correlations found
for local galaxies ranging over more than three orders of magnitude in
far--IR luminosity.

{\em Figure 3c}: Based on data from Kohno et al.\ (\citeyear{koh01})
and Kohno (\citeyear{koh05}), Imanishi et al.\ (\citeyear{ima06})
argue that AGN--dominated galaxies have higher HCN/HCO$^+$ and HCN/CO
ratios than starburst--dominated galaxies.  In this diagram,
AGN--dominated galaxies thus fall on the upper right side, and
starburst--dominated galaxies fall on the lower left side.  For the
local starburst/spiral sample, we have taken the $L'_{\rm
  HCN}$/$L'_{\rm HCO^+}$ ratios from Nguyen-Q-Rieu et al.\
(\citeyear{ngu92}) and the $L'_{\rm HCN}$/$L'_{\rm CO}$ ratios from
Gao \& Solomon (\citeyear{gao04a}).  The Cloverleaf clearly falls on
the lower left side of the diagram, putting it in the region of
'starburst--dominated' galaxies. Indeed, by decomposition of the dust
spectrum into a warm (115\,K) and a cooler (50\,K) component, Weiss et
al.\ (\citeyear{wei03}) find that about 60\% of the dust emission
emerges from the cooler component, which may well be dominated by
heating from star formation. However, based on an Arp\,220 template,
Solomon et al.\ (\citeyear{sol03}) have shown that only about 20\% of
the total far--IR luminosity is powered by the starburst.  It is
therefore unclear whether or not this diagram should indeed be used to
constrain the properties of high--$z$ quasars.

{\em Figure 3d}: Based on their recent study of local (U)LIRGs,
Graci\'a--Carpio et al.\ (\citeyear{gra06}) suggest that the $L'_{\rm
  HCN}$/$L'_{\rm HCO^+}$ ratio correlates with $L_{\rm FIR}$; their
results indicate that HCN may not be an unbiased tracer of star
formation. We show this relation including their sample together with
the spiral/starburst sample described above and the Cloverleaf. We
find no evidence for any correlation between $L'_{\rm HCN}$/$L'_{\rm
  HCO^+}$ and $L_{\rm FIR}$ over the increased luminosity and redshift
range. We note that if the relation suggested by Graci\'a--Carpio et
al.\ (\citeyear{gra06}) were to hold, we would not have been able to
detect the \ahco\ line in the Cloverleaf, as $L'_{\rm HCN}$/$L'_{\rm
  HCO^+}$ would be $\sim$3.

As discussed above, the $L'_{\rm HCN}$/$L'_{\rm HCO^+}$ ratio is
consistent with unity over a large range in far--IR luminosities.
Together with multi--transition studies of both molecules available
for some of the local galaxies (e.g.\ Seaquist \& Frayer
\citeyear{sea00}), this result suggests that HCO$^+$ does not require
special conditions to be excited.  Also, it indicates that HCO$^+$ and
HCN trace physically similar regions.  Although being consistent with
unity within the error bars, the ratio of HCN and HCO$^+$ luminosities
may be larger than 1 in the Cloverleaf, and several mechanisms to
explain such a difference have been discussed in the literature.  It
has been argued that HCN emission in molecular clouds can be enhanced
by mid--IR pumping of a 14\,$\mu$m vibrational band (Aalto et al.\
\citeyear{aal95}), but HCO$^+$ can be mid--IR--pumped under very
similar conditions via a 12\,$\mu$m vibrational band (Graci\'a--Carpio
et al.\ \citeyear{gra06}). Also, Gao \& Solomon (\citeyear{gao04a})
have found that this mechanism does not appear to play a major role
(their Fig.~5).  It has also been suggested that the chemical
abundance of HCN can be enhanced relative to CO and HCO$^+$ in the
ambient X--ray radiation field of a strong AGN (Lepp \& Dalgarno
\citeyear{lep96}).  It is also possible that the HCO$^+$ abundance is
decreased due to the ionizing field produced by cosmic rays (Seaquist
\& Frayer \citeyear{sea00}): Cosmic rays ionize H$_2$, leading to the
production of H$_3^+$, which reacts with CO to form HCO$^+$. The
abundance of HCO$^+$ is thus affected by ratio of cosmic--ray
ionization rate and gas density.  While a higher ionizing flux favours
the production of HCO$^+$, it also increases the number of free
electrons, which leads to a higher probability for recombination: At a
gas density of $3\times10^4$\,cm$^{-3}$, an ionizing field comparable
to that of the Galaxy may already be strong enough to significantly
decrease the abundance of HCO$^+$ due to dissociative recombination of
H$_3^+$ (e.g.\ Phillips \& Lazio \citeyear{phi95}), and the ionizing
field in the Cloverleaf is probably much stronger than the Galactic
one.

Our finding that the HCN--to--HCO$^+$ luminosity ratio in the
Cloverleaf is close to unity implies that the processes discussed here
likely do not play a dominant role.  All these chemical arguments,
assume that the HCN and HCO$^+$ $J=1\to0$ opacities are low and that
the observed line intensities scale with the underlying molecular
abundances. Observations of the $^{13}$C baring isotophomeres of
HCO$^+$ and HCN in nearby starburst galaxies have shown that the
HCO$^+$/H$^{13}$CO$^+$ and HCN/H$^{13}$CN ratios are similar to those
seen in CO/$^{13}$CO (Nguyen-Q-Rieu et al.\ \citeyear{ngu92}; Wang et
al.\ \citeyear{wan04}). From this it has been concluded that that the
HCO$^+$ and HCN opacities are similar to those in CO with
$\tau\simeq3-4$ (see also Henkel et al.\ \citeyear{hen93}), i.e.\ the
emission is optically thick.  This would imply that $L'_{\rm
  HCN}$/$L'_{\rm HCO^+}$ is solely determined by the relative area
filling factors and excitation temperatures of both molecules. As the
\ahco\ and \ahcn\ excitation is likely to be close to thermalized for
densities of $n({\rm H_2}) \gtrsim 10^5\,$cm$^{-3}$, this would
naturally explain $L'_{\rm HCN}$/$L'_{\rm HCO^+} \simeq 1$ assuming
that both molecules trace regions of similar density (i.e., similar
volume).  We conclude that HCO$^+$ compares favourably with HCN in
terms of being a good tracer for dense molecular gas even in
FIR--bright objects at high redshift.

\acknowledgments 
The National Radio Astronomy Observatory is operated by Associated
Universities Inc., under cooperative agreement with the National
Science Foundation.  D.~ R.\ acknowledges support from the Deutsche
Forschungsgemeinschaft (DFG) Priority Programme 1177.  C.~C.\
acknowledges support from the Max-Planck-Gesellschaft and the
Alexander von Humboldt-Stiftung through the
Max-Planck-Forschungspreis.  The authors would like to thank the
referee for useful comments which helped to improve the manuscript.



\begin{figure}
\epsscale{.5}
\plotone{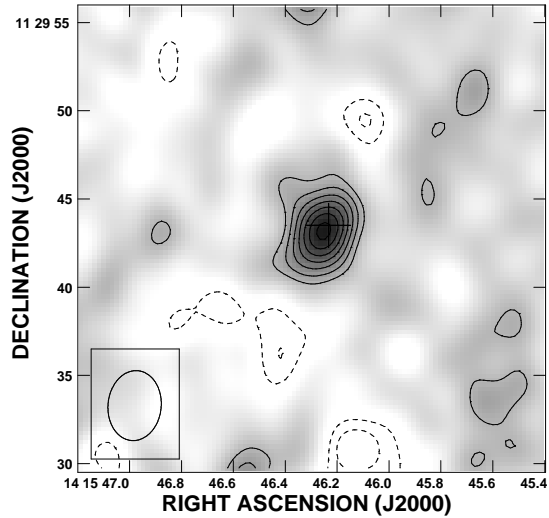}
\caption{VLA detection of \ahco\ towards the Cloverleaf quasar at a 
resolution of 4.0\,$''$$\times$3.0\,$''$ (as indicated in the bottom 
left corner). Continuum emission was subtracted. The source is marginally 
resolved. The cross indicates the geometrical center of the CO emission 
in the Cloverleaf (Alloin et al.\ \citeyear{all97}, see text). This 
continuum--subtracted map is integrated over the central 411\kms\ 
(34.375\,MHz). Contours are shown at 
(-3, -2, 2, 3, 4, 5, 6, 7, 8)$\times\sigma$ 
(1$\sigma = 16\,\mu$Jy beam$^{-1}$). \label{f1}}
\vspace{4.5mm}

\end{figure}


\begin{figure}
\epsscale{1.0}
\plotone{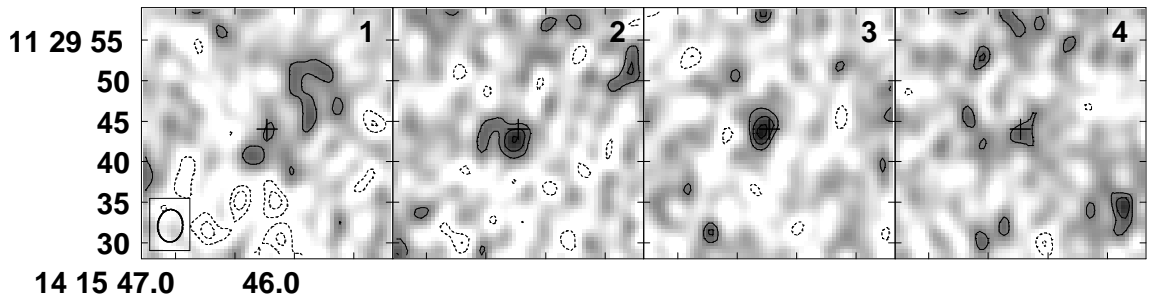}
\caption{Channel maps of the \ahco\ emission (same region is shown as 
  in Fig.~1). One channel width is 6.25\,MHz, or 75\,km\,s$^{-1}$
  (frequencies increase with channel number and are shown at
  25057.228, 25063.478, 25069.728 and 25075.978\,MHz).  Contours are
  shown at (-3, -2, 2, 3, 4)$\times \sigma$
  (1$\sigma = 50\,\mu$Jy\,beam$^{-1}$). The beam size 
  (4.0\,$''$$\times$3.0\,$''$) is shown in the bottom left corner; the 
  cross indicates the same position as in Fig.\ 1. \label{f2}}
\end{figure}


\begin{figure}
\epsscale{1.0}
\plotone{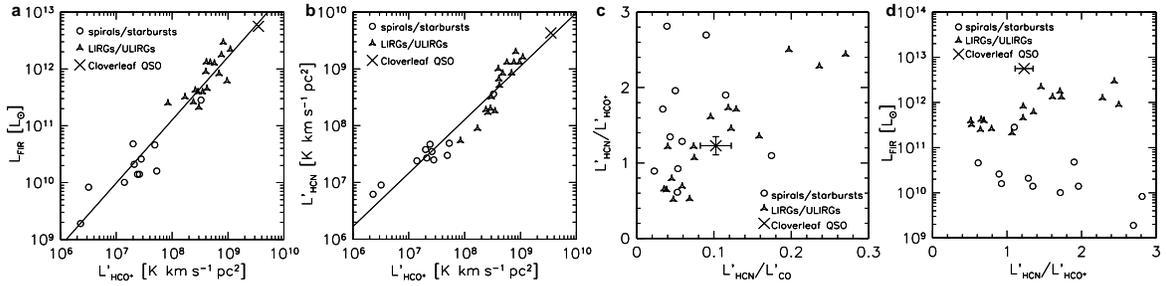}
\caption{HCO$^+$ luminosity relations for a sample of low--$z$ spiral and 
starburst galaxies (Nguyen-Q-Rieu et al.\ \citeyear{ngu92}; Imanishi et 
al.\ \citeyear{ima04}; Gao \& Solomon \citeyear{gao04a}), low--$z$ (U)LIRGs 
(Graci\'a--Carpio et al.\ \citeyear{gra06}; Imanishi et al.\ 
\citeyear{ima06}), and the Cloverleaf (this work; Solomon et al.\ 
\citeyear{sol03}; Weiss et al.\ \citeyear{wei03}). The Cloverleaf 
luminosities are corrected for gravitational lensing ($\mu_{\rm L} = 11$). 
The solid lines are least squares fits to all data except the Cloverleaf. 
The error bars shown for the Cloverleaf indicate the statistical errors of 
the line luminosity measurements. See text for more details. \label{f3}}
\end{figure}


\begin{deluxetable}{ l c c c }
\tabletypesize{\scriptsize}
\tablecaption{CO, HCN, and HCO$^+$ line luminosities in the Cloverleaf. \label{tab-1}}
\tablehead{
& $S_{\nu}$ & $L'$ & Ref. \\
& [$\mu$Jy] & [10$^9$\,K\,\kms\,pc$^2$] & }
\startdata
\ahco\ & 193 $\pm$ 28 & 3.5 $\pm$ 0.3 & 1 \\
\ahcn\ & 240 $\pm$ 40 & 4.3 $\pm$ 0.5 & 2 \\
\cco\ & 30000 $\pm$ 1700 & 42 $\pm$ 7 & 3 \\ 
\vspace{-2mm}
\enddata 
\tablerefs{${}$[1] This work, [2] Solomon \etal\ (\citeyear{sol03}), [3] Weiss 
\etal\ (\citeyear{wei03}).}
\tablecomments{${}$Luminosities are derived as described by Solomon \etal\ 
(\citeyear{sol92b}): 
$L'[{\rm K \ts km\ts s^{-1} pc^2}] = 3.25 \times 10^7 \times I \times \nu_{\rm obs}^{-2} \times D_{\rm L}^2 \times (1+z)^{-3}$, where 
$I$ is the velocity--integrated line flux in Jy \kms, $D_{\rm L}$ 
is the luminosity distance in Mpc ($z=2.55784$, Weiss \etal\ 
\citeyear{wei03}), and $\nu_{\rm obs}$ is the observed frequency in GHz.
All given luminosities are corrected for this lensing maginification. The 
HCN and HCO$^+$ luminosities are corrected for the finite source size 
relative to the synthesized VLA beam (see text). 
} 
\end{deluxetable}



\begin{thebibliography}{}

\bibitem[Aalto et al.(1995)]{aal95} Aalto, S., Booth, R.~S., Black,
  J.~H., \& Johansson, L.~E.~B.\ 1995, A\&A, 300, 369
\bibitem[Alloin et al.(1997)]{all97} Alloin, D., Guilloteau, S.,
  Barvainis, R., Antonucci, R., \& Tacconi, L., 1997, A\&A, 321, 24
\bibitem[Brouillet et al.(2005)]{bro05} Brouillet, N., Muller, S.,
  Herpin, F., \etal\ 2005, A\&A, 429, 153
\bibitem[Carilli et al.(2005)]{car04} Carilli, C.~L., Solomon, P.~M.,
  Vanden Bout, P.~A., \etal\ 2005, ApJ, 618, 586
\bibitem[Gao \& Solomon(2004a)]{gao04a} Gao, Y., \& Solomon, P.~M.\
  2004a, ApJS, 152, 63
\bibitem[Gao \& Solomon(2004b)]{gao04b} Gao, Y., \& Solomon, P.~M.\
  2004b, ApJ, 606, 271
\bibitem[Graci\'a--Carpio et al.(2006)]{gra06} Graci\'a-Carpio, J.,
  Garc\'ia-Burillo, S., Planesas, P., \& Colina, L.\ 2006, ApJ, 640,
  L135
\bibitem[Henkel et al.(1993)]{hen93} Henkel, C., Mauersberger, R.,
  Wiklind, T., \etal\ 1993, A\&A, 268, L17
\bibitem[Imanishi et al.(2004)]{ima04} Imanishi, M., Nakanishi, K.,
  Kuno, N., \& Kohno, K.\ 2004, AJ, 128, 2037
\bibitem[Imanishi et al.(2006)]{ima06} Imanishi, M., Nakanishi, K., \&
  Kohno, K.\ 2006, AJ, in press
\bibitem[Kohno et al.(2001)]{koh01} Kohno, K., Matsushita, S.,
  Vila-Vilaro, B., \etal\ 2001, ASP Conf.\ Ser.\ 249: The Central kpc
  of Starbursts and AGN, 672
\bibitem[Kohno(2005)]{koh05} Kohno, K.\ 2005, AIP Conf.\ Ser.\ 783:
  The Evolution of Starbursts, 203
\bibitem[Lepp \& Dalgarno(1996)]{lep96} Lepp, S., \& Dalgarno, A.\
  1996, A\&A, 306, L21
\bibitem[Nguyen-Q-Rieu et al.(1992)]{ngu92} Nguyen-Q-Rieu, Jackson,
  J.~M., Henkel, C., Truong-Bach, \& Mauersberger, R.\ 1992, ApJ, 399,
  521
\bibitem[Phillips \& Lazio(1995)]{phi95} Phillips, J.~A., \& Lazio,
  T.~J.~W.\ 1995, ApJ, 442, L37
\bibitem[Seaquist \& Frayer(2000)]{sea00} Seaquist, E.~R., \& Frayer,
  D.~T.\ 2000, ApJ, 540, 765
\bibitem[Solomon et al.(1992a)]{sol92a} Solomon, P., Downes, D., \&
  Radford, S.\ 1992a, ApJ, 387, L55
\bibitem[Solomon et al.(1992b)]{sol92b} Solomon, P.~M., Radford,
  S.~J.~E., \& Downes, D.\ 1992b, Nature, 356, 318
\bibitem[Solomon et al.(2003)]{sol03} Solomon, P., Vanden Bout, P.,
  Carilli, C., \& Guelin, M.\ 2003, Nature 426, 636
\bibitem[Solomon \& Vanden Bout(2005)]{sv05} Solomon, P.~M., \& Vanden
  Bout, P.~A.\ 2005, ARA\&A, 43, 677
\bibitem[Spergel et al.(2003)]{spe03} Spergel, D.~N., Verde, L.,
  Peiris, H.~V., \etal\ 2003, ApJS, 148, 175
\bibitem[Spergel et al.(2006)]{spe06} Spergel, D.~N., Bean, R.,
  Dor\'e, O., \etal\ 2006, ApJ, submitted
\bibitem[Tsuboi et al.(1999)]{tsu99} Tsuboi, M., Miyazaki, A.,
  Imaizumi, S., \& Nakai, N.\ 1999, PASJ, 51, 479
\bibitem[Usero et al.(2004)]{use04} Usero, A., Garc\'ia-Burillo, S.,
  Fuente, A., \etal\ 2004, A\&A, 419, 897
\bibitem[Vanden Bout et al.(2004)]{vdb04} Vanden Bout, P., Solomon,
  P., \& Maddalena, R.\ 2004, ApJ, 614, L97
\bibitem[Venturini \& Solomon(2003)]{ven03} Venturini, S., \& Solomon,
  P.~M.\ 2003, ApJ, 590, 740
\bibitem[Wagg et al.(2005)]{wag05} Wagg, J., Wilner, D.~J., Neri, R.,
  Downes, D., \& Wiklind, T.\ 2005, ApJ, 634, L13
\bibitem[Wang et al.(2004)]{wan04} Wang, M., Henkel, C., Chin, Y.-N.,
  \etal\ 2004, A\&A, 422, 883
\bibitem[Weiss et al.(2003)]{wei03} Wei\ss, A., Henkel, C., Downes,
  D., \& Walter, F.\ 2003, A\&A, 409, L41
\bibitem[Weiss et al.(2005)]{wei05} Wei\ss, A., Downes, D., Henkel,
  C., \& Walter, F.\ 2005, A\&A, 429, L25
\bibitem[Wilner et al.(1995)]{wil95} Wilner, D.~J., Zhao, J.-H., \&
  Ho, P.~T.~P.\ 1995, ApJ, 453, L91

\end{thebibliography}
\end{document}